\documentstyle[dcolumn]{l-aa}

\def\min{\raisebox{0.4ex}{$\scriptscriptstyle{-}$}}
\def\plu{\raisebox{0.4ex}{$\scriptscriptstyle{+}$}}
\def\plumi{\raisebox{0.4ex}{$\scriptscriptstyle{\pm}$}}

\def\et{{\it et\thinspace al.}\ }                 
\def\kms{km\thinspace s$^{-1}$ }                   
\def\kmx{km\thinspace s$^{-1}$}                   
\def\deg{\ifmmode^\circ\else$^\circ$\fi}          
\def\arcs{\ifmmode {'' }\else $'' $\fi}           
\def\arcm{\ifmmode {' }\else $' $\fi}             
\def\buildrel#1\over#2{\mathrel{\mathop{\null#2}\limits^{#1}}}
\def\mper{\ifmmode \buildrel m\over . \else $\buildrel m\over .$\fi}
\def\hper{\ifmmode \rlap.^{h}\else $\rlap{.}^h$\fi}    
\def\sper{\ifmmode \rlap.^{s}\else $\rlap{.}^s$\fi}    
\def\arcsper{\ifmmode \rlap.{' }\else $\rlap{.}' $\fi} 
\def\arcmper{\ifmmode \rlap.{'' }\else $\rlap{.}'' $\fi}
\def\gapprox{\ifmmode {_ >\atop{^\sim}}\else $_ >\atop{^\sim}$\fi} 
\def\lapprox{\ifmmode {_ <\atop{^\sim}}\else $_ <\atop{^\sim}$\fi} 
\def\onerule{\noalign{\medskip\hrule\medskip}}        

\begin{document}

\thesaurus{03        
           (04.03.1; 
            11.03.1; 
            11.04.1) 
}

\title{Structure and kinematics of galaxy clusters}

\subtitle{II. Substructures and luminosity segregation}

\author{P.~Stein \inst{1,} \inst{2}}

\offprints{P.~Stein (Universitat de Barcelona)}

\institute{Astronomisches Institut der Universit\"at Basel, 
Venusstrasse 7, CH--4102 Binningen, Switzerland
\and Departament d'Astronomia i Meteorologia, Universitat de Barcelona,
Avenida Diagonal 647, E--08028 Barcelona, Spain}

\date{accepted June 14, 1996}

\maketitle

\begin{abstract}
A homogeneous sample of galaxy redshifts in the core regions (R $\leq$
0.5 h$^{-1}$ Mpc) of 12 clusters is used to measure the frequency of
substructure with different tests. In 50 \% of the cases substructure
is detected, a frequency which agrees well with previous studies of
cluster cores. Magnitude information and rough morphological classes
are also available for 80 \% of the sample, allowing us to confirm that
bright galaxies ($M \lapprox$ -22$^{mag}$) have a significantly lower
velocity dispersion than the rest. Elliptical galaxies are the main
responsibles for this luminosity segregation in velocity space, whereas
no segregation can be seen for spiral galaxies. Given the coincidence
of substructure and luminosity segregation, a cluster model with an old
population of ellipticals which are under the effect of dynamical
friction in each subcluster is thus favoured by these observations. 
Spiral galaxies seem to be late arrivals, or are passing in front or
behind the core of the cluster.

\end{abstract}

\input boxedeps
\EmulateRokicki
\SetepsfEPSFSpecial

\section{Introduction}
One of the main results of recent galaxy cluster research is the
unambiguous finding that these high-density structures are still
continuing to be built or at least significantly reshaped.
Undoubtedly, the discovery of substructures has
played an important role, changing our understanding about the degree
of dynamical evolution in galaxy clusters.\\ Several statistical
treatments to measure the frequency of substructures have been
presented during the last decade. Nevertheless, the question about the
significance of substructure detections on small scales or in the {\it
core} regions of clusters has been touched only in a few works
(Fitchett \& Webster 1987; Mellier \et 1988; Escalera \et 1992; 
Salvador--Sol\'e \et 1993).\\ 
It is interesting 
that the frequency and degree of clumpiness in the centers of
clusters could be helpful in establishing the density profile of dark
matter, and even allow an estimation of $\Omega$ (e.g. Richstone \et
1992).  If namely dark matter was strongly concentrated towards the
cluster center one would expect tidal forces acting towards the
disruption of subclumps (see Gonz\'alez-Casado \et 1994 for a
discussion).\\ Examining the very centers of galaxy clusters is
rewarding for another reason. 
The importance of accretion or ``cannibalism'' (Ostriker \&
Hausman 1977) during the formation of a cluster is not well
established, because right in the cores of galaxy clusters the galaxy
velocity dispersion is expected to be too high to allow an efficient
dynamical friction (Merrifield \& Kent 1989, Gebhardt \& Beers 1991,
Blakeslee \& Tonry 1992). An attractive solution to this problem has
been proposed by Merritt (1985).\\
As also numerical simulations suggest (West \& Richstone
1988, Serna \et 1994), dynamical friction could be an important
mechanism during the evolution of galaxy clusters. If this is the case,
it should be possible to detect signs of equipartition of kinetic
energies for the most massive galaxies in dense regions, leading to a
correlation of velocity dispersion with galaxy luminosity. Early
observational studies of mass segregation were mostly limited to the
positions of galaxies in projection, with the exception of e.g.
Chincarini \& Rood (1977), and no general agreement was foud about the
relevance of luminosity segregation in velocity space.  Recently,
Biviano \et (1992) once again stated the evidence for mass segregation
in the {\it velocity distribution} of a merged cluster sample taken
from the literature. Presumably, the phenomenon is easily overlooked in
individual clusters studies because of the limited number of galaxies
involved. \\ On the other side, a difference in the kinematic
properties between early and late--type galaxies has also been detected
(Binggeli \et 1987, Sodr\'e \et 1989). This means that one should look
separately at early and late--type galaxies, because galaxies which
didn't take part in most of the cluster evolution are
not expected to show signs of mass segregation.\\ This paper addresses 
the problem of mass segregation and central substructure
based on a homogeneous sample of kinematical data (Stein 1996,
hereafter Paper I). It is organized as follows: Section 2 gives
a description of the data used, i.e. the source of redshifts,
magnitudes and types for the galaxies, as well as cluster definition
and selection. The analysis of substructures on
individual clusters is done in Section 3, while luminosity segregation
is measured on the merged cluster sample (Section 4). The results are
then discussed in Section 5.

\section{The sample}
The present work is based on a redshift catalogue involving the central
regions of 15 clusters of different richnesses in the redshift range
0.01 $\lapprox$ z $\lapprox$ 0.06 (Paper I).  Note that several of the
clusters are well studied, nearby objects like Centaurus and Abell 400
or have been extensively observed for specific purposes (e.g. Abell
3558~: Bardelli \et 1994). Inside these regions
nearly complete \mbox{(80 \%)} redshift information down to M$_B$ $\approx$
18$^{mag}$ could be obtained, as well as raw morphological types for most
of the galaxies. The present sample differs from those of
previous investigations because it concentrates onto the very cores of
the clusters (R $<$ 0.5 h$^{-1}$ Mpc), and because of the homogeneity of
kinematical information. \\ 
Cluster centers were chosen in order of preference from (a) published X-ray
centers (Lahav \et 1989, Edge \& Stewart 1991), (b) the position of the
cd--galaxy, if present, (c) the Abell cluster center or (d) by taking
the center of the relevant Optopus field (see Paper I), which roughly 
coincides with the galaxy density peak. The only exception to this
procedure was Abell 1736, whose OPTOPUS field was
offset by 0.05 h$^{-1}$ Mpc with respect to the X-ray center.

A common selection radius of 0.5 h$^{-1}$ Mpc around each cluster
centre was therefore chosen, using approximate mean cluster velocities
and assuming all clusters to be located in the Hubble flow. For Hydra
and Centaurus (z~$\approx$~0.01) two adjacent fields had been taken,
and for the other very nearby clusters data from the literature was
available to help fill the region.

Table 1 lists the fields and also gives the name of each
cluster around whose centre the selection has been done (2), 
kind of center determination (3), where X : X--rays, cd :
cd--galaxy position, A : Abell center, O : center of Optopus
field, as well as centre coordinates (4), and selection radius in
arcminutes (5). 

\setcounter{table}{0}

\begin{table*}[htbp]
\caption{Cluster sample}
\begin{tabular}[l]{rp{4.0cm}cllr}
\onerule
\multicolumn{1}{c}{Field} & 
\multicolumn{1}{c}{Name(s)} & 
\multicolumn{1}{c}{C} & 
\multicolumn{2}{c}{$\alpha$ (B1950) $\delta$}& 
\multicolumn{1}{c}{R}\\

\multicolumn{1}{c}{(1)} & 
\multicolumn{1}{c}{(2)}& 
\multicolumn{1}{c}{(3)}& 
\multicolumn{2}{c}{(4)} & 
\multicolumn{1}{c}{(5)} \\
\onerule                    
 1&Abell S0301, \mbox{DC~0247\min31}&cd& 02~47~27& \min31~23~44 & 26\arcm  \\ 
 2&Abell 0400                       & X& 02~55~03& \plu05~49~32 & 26\arcm  \\
 3&Abell 1016                       &cd& 10~24~28& \plu11~15~56 & 19\arcm  \\
 4&Abell 1060, Hydra                & X& 10~34~22& \min27~15~58 & 45\arcm  \\
 5&Abell S0639                      & O& 10~37~48& \min46~01    & 29\arcm  \\
 6&Abell 3526, \mbox{Centaurus}     & X& 12~46~03& \min41~02~28 & 48\arcm  \\
 7&Abell S0721                      & A& 13~03~18& \min37~19~00 & 13\arcm  \\
 8&Abell 3556                       & O& 13~20~24& \min31~27    & 13\arcm  \\
 9&Abell 1736,  \mbox{DC~1324\min27}& X& 13~24~46& \min26~55~24 & 18\arcm  \\
10&CL 1322\min30                    & O& 13~22~00& \min30~02    & 42\arcm  \\
11&Abell 3558, \mbox{Shapley~8}     & X& 13~25~08& \min31~14~13 & 13\arcm  \\
12&Abell S805,  \mbox{DC~1842\min63}&cd& 18~42~35& \min63~23~04 & 40\arcm  \\
13&Abell 3733                       &cd& 20~58~39& \min28~15~22 & 16\arcm  \\
14&Abell 3880                       &cd& 22~25~05& \min30~49~52 & 11\arcm  \\
15&Abell 4038,  \mbox{Klemola~44}   & X& 23~45~18& \min28~26~00 & 21\arcm  \\
\onerule
\end{tabular}          
\end{table*}

\subsection{Membership selection}
As a first step towards meaningful cluster definitions, we then looked
at velocity histograms in the direction of all 15 fields, eliminating
obvious (5 $\sigma$) fore-- or background galaxies. In one case (field
12) two distinct structures could be identified~: the main cluster at 
v~=~4500 \kms and a smaller group at v~=~11000 \kmx. 
Moreover, some non--gaussian velocity histograms with indication of
bimodality could be recognized. A statistical test was thus
applied which returns the likelihood that the biggest gap in the
dataset could occur in a normal distribution (adapted from the 
ROSTAT package, Beers \et 1990). 
Only in the case of field 9 a gap was found whose size was
inconsistent at the 5 \% level with an underlying gaussian
distribution. Field 9 was thus considered bimodal and subdivided at
v~=~11500 \kmx, the big gap location.
To assess final cluster membership, a 3--$\sigma$ clipping technique
(Yahil \& Vidal 1977) 
was then applied to each of the 17 galaxy units recognized so far. 
Table 2 lists the
number of member galaxies after 3--$\sigma$ clipping (column 3), as
well as the resulting limits in redshift space (4). 

Mean cluster velocities and velocity dispersions were then determined
for each cluster in the sample, using biweight estimators (Beers \et
1990). These estimators have the advantage of being robust against
outliers and are particularly useful when working with small datasets.
Cosmological effects are taken into consideration following Danese \et
(1980). The resulting values and their uncertainties are listed in Table
2, columns 5 and 6.

Our cluster mean velocities and
velocity dispersions do agree quite well with those quoted by
Struble \& Rood (1991) and Girardi \et (1993). Nevertheless, there are
some cases where a discrepancy is found. The reason
herefore must be related to the effect of substructures in the
central cluster regions, which tend to inflate velocity dispersions, 
and to the presence of luminous galaxies of low
dispersion in the cluster cores (as shown in
the next sections). Nevertheless, given that our results are
virtually free from larger scale contamination, we consider them good
estimators of central mean velocity and velocity dispersion. With the
present data it was not possible to discern the bimodal structure of
Centaurus using a gapper test. For this reason we
give a global value for the mean cluster velocity and for the velocity
dispersion. Nevertheless, substructure is detected in this and in some
of the other clusters, as we will see below. It may be, therefore, that
the kinematic parameters given in Table 2 do not reflect the true
dynamical state in 50 \% of the cases. They should be treated as
first estimations and were calculated because a measure for the cluster
dispersion is needed for the Lee--test simulations (see below).

Some of the galaxy samples are clearly too poor to be used for the
substructure analysis and have to be discarded. Computation of the
Dressler \& Shectman (1988) substructure test requires the evaluation
of velocity dispersions from samples of $\sqrt{N}$ neighbours around each
galaxy, where $N$ is the total number of galaxies (Bird 1994). Given that
5 is a lower limit for the determination of standard deviations, we
chose a number of $N \geq 5^2 = 25$ galaxies as the minimum richness.

We are thus left with 12 clusters (flag ``Y'' in Table 2, column 7)
and a total of 576 galaxy redshifts, 2/3 of which are taken from Paper
I and have mean errors well below 50 km~s$^{-1}$.

\begin{table*}[htbp]
\caption{Redshift ranges and kinematical parameters}
\begin{tabular}[l]{rp{4.6cm}rD{!}{-}{-1}D{!}{\plumi}{-1}
D{!}{\plumi}{-1}c}
\onerule
\multicolumn{1}{c}{Field} & 
\multicolumn{1}{c}{Name(s)} & 
\multicolumn{1}{c}{$\#$} & 
\multicolumn{1}{c}{cz--range}&
\multicolumn{1}{c}{$\langle v\rangle\plumi\Delta v$} & 
\multicolumn{1}{c}{$\sigma\plumi\Delta\sigma$} &
\multicolumn{1}{c}{inclusion} 
\\
\multicolumn{3}{c}{} & 
\multicolumn{1}{c}{[\kmx]}& 
\multicolumn{1}{c}{[\kmx]}& 
\multicolumn{1}{c}{[\kmx]}& 
\multicolumn{1}{c}{flag}\\

\multicolumn{1}{c}{(1)} & 
\multicolumn{1}{c}{(2)}& 
\multicolumn{1}{c}{(3)}& 
\multicolumn{1}{c}{(4)} & 
\multicolumn{1}{c}{(5)} & 
\multicolumn{1}{c}{(6)} & 
\multicolumn{1}{c}{(7)}
\\
\onerule                             
   1& Abell S0301, \mbox{DC~0247\min31}& 25&  5500! 8000 &  6867! 98 & 473!99 & Y \\ 
   2& Abell 0400                       & 73&  5500! 8500 &  7057! 65 & 547!57 & Y \\
   3& Abell 1016                       & 22&  9000!10500 &  9646! 56 & 252!51 & \\          
   4& Abell 1060, Hydra                & 76&  2000! 6000 &  3867! 84 & 727!69 & Y \\
   5& Abell S0639                      & 32&  5500! 7500 &  6194! 78 & 431!52 & Y \\
   6& Abell 3526, \mbox{Centaurus}     & 64&  1000! 6000 &  3688!120 & 952!63 & Y \\
   7& Abell S0721                      & 29& 13500!16500 & 14936!134 & 703!70 & Y \\
   8& Abell 3556                       & 30& 13500!15500 & 14574! 86 & 459!44 & Y \\
  9a&                                  & 16&  9500!11500 & 10537!102 & 385!60 & \\          
  9b& Abell 1736,  \mbox{DC~1324\min27}& 48& 11000!16000 & 13734!130 & 889!81 & Y \\        
  10& CL 1322\min30                    & 18&  3500! 5000 &  4222! 70 & 278!52 & \\          
  11& Abell 3558, \mbox{Shapley~8}     & 59& 11000!18000 & 14242!155 &1183!100& Y \\
 12a& Abell S805,  \mbox{DC~1842\min63}& 54&  3000! 6500 &  4603! 87 & 621!64 & Y \\
 12b&                                  & 11& 10000!11500 & 10771!121 & 367!51 & \\          
  13& Abell 3733                       & 27& 10000!13000 & 11716!103 & 522!84 & Y \\
  14& Abell 3880                       & 22& 15000!19000 & 17513!188 & 855!148& \\          
  15& Abell 4038,  \mbox{Klemola~44}   & 59&  7000!11000 &  8936!118 & 896!66 & Y \\
\onerule
\end{tabular}          
\end{table*}

\subsection{Photometric data}
As we wanted to look at dynamical friction
effects, luminosity information was necessary. For three of the
clusters detailed photometrical studies could already be found in the
literature~: A 3526 = Centaurus by Dickens \et (1986), A~S 805 by
Millington \& Peach (1989) and \mbox{A 4038} by Green \et (1990). In addition,
b$_j$ magnitudes from the COSMOS catalogue were kindly provided by
H.~MacGillivray (1993) for ten clusters of high galactic latitude.
Given the selection of high latitude clusters no correction for
reddening or extinction was applied.\\ A comparison of COSMOS
magnitudes with those of Green \et (1990) in A 4038 shows an excellent
agreement, with differences of only a few hundredths magnitudes. For
some of the brightest spiral galaxies COSMOS magnitudes were not
available (due to the problem of resolved HII regions), which meant
that magnitudes had to be taken from a bright galaxy catalogue.
It should be noted that magnitudes from the COSMOS
catalogue are b$_j$ magnitudes, while those for Centaurus are G26.5
magnitudes, and for most of the other clusters generic optical
magnitudes were taken from different sources.  Magnitudes were not
transformed to a common scale, because for our statistical analysis a
spread of a few tenth of magnitudes for individual galaxies could be
taken into account.

\subsection{Morphological data}
Galaxy types were searched for in the literature, mainly resulting in a
sample from Dressler's (1980) and UGC (Nilson 1973) catalogues, as well
as from Huchra's (1991) collection, as implemented in the DIRA2
database (Astronet Data Base Group Italy, Bologna).  All galaxies were
then divided into three classes: E, S0 and S. In addition, many of the
remaining galaxies were classified by the author into one of the above
classes, after visual inspection of ESO--Schmidt plates. Because
careful classification is difficult on this kind of plates, a check was
made involving 105 galaxies with independently known types. Of these,
15 were classified by the author with an ``earlier'' type and 13 with a
``later'' type than literature values, corresponding to an agreement of
around 75 \%.

\section{Substructures}
Ideally, substructure tests should be applied to galaxy samples which
are complete in magnitude. Here, we will use all the available data
without regard to the magnitudes, for different reasons. First, the
small sample size makes any further restriction
unreasonable. In addition, complete magnitude information is not
available for all cluster fields. Nevertheless, in those fields where
magnitudes were available from COSMOS, redshift completeness amounts to 
ca. \mbox{75 \%} of the galaxies down to 18$^{mag}$.  Given 
the small field of
view and the extensive redshift coverage, we will
concentrate on substructure analysis methods which make use of the
velocity information. We chose the test of Dressler \& Shectman (1988,
hereafter DS--test), the Lee test (Fitchett 1988), and several tests
that check for departures from normality of the galaxy velocity
distribution. Note that:\\ a) For the DS--test we followed the same
procedure as Bird (1994), using $\sqrt{N}$ galaxies to define the
neighbourhood of each galaxy, where $N$ is the total number of galaxies
involved in the analysis.\\ b) To assess the significance of
substructure detections for the Lee test we performed the same test on
100 cluster simulations.  Modelling of the simulated clusters was done
with an 1/r galaxy density profile (Fitchett \& Webster 1987) 
and with the same velocity dispersion as in
Table 2. Steps of 4 degrees have been used for the
orientation of the projection axis in the X-Y-cz plane. Note that the
simulations for field 9b were done with an offset of 0.05 h$^{-1}$ Mpc
between the peak of the 1/r density distribution and the center of the
circular selection region. A higher likelihood for the presence of
substructure would result if the offset was not accounted for (92 \%
instead of 87 \%).\\ c) Three different tests for normality have been
employed, i.e. the U$^2$, W$^2$ and A$^2$ tests taken from the ROSTAT
package (Beers \et 1990), the resulting significance being the mean
from these three tests, which agree very well.\\ Each of these
substructure 
tests has its strengths and weaknesses. For example, the DS--test
measures the deviation of the local from the global kinematics,
allowing to detect efficiently small offcenter groups, but sometimes
failing in cases where two clumps of equal size and different mean
velocities overlap in projection. On the other side, the Lee test is
designed to be a very general maximum--likelihood method, but
computational constraints confine its use to detecting bimodal
structures of comparable size, thus being unable to state about the presence
or absence of multimodal structures. It is also trivial that testing
the gaussianity of a velocity sample alone cannot give clues about
substructures which have same means and dispersions but differing
locations in the plane of the sky.  We see that all of these tests are
bound to miss some manifestations of substructure, each one being
sensitive to some particular configuration. For these reasons, a
combination of all methods should allow a better judgement to be
made. Our statement about the existence of substructure relies upon the
fact that at least one of the three detection methods could find
significant signs for it. Results are shown in Table 3, which gives
field number (column 1), name of the cluster (2), significance
level for the tests mentioned above (3--5), and total significance
level for substructure (6), which is the highest value of columns 3,
4, and 5.
As can be seen from the analysis results, 50 \% of the clusters show
clear signs of substructure (5 \% significance level).  This frequency
of substructure in cluster cores has to be considered a lower level
because of the intrinsic property of individual tests to miss some
manifestations of substructure.  It should be noted that the same order
of magnitude for substructure frequency has been found by other authors
(e.g. Escalera \et 1994; West 1994), in particular also 
involving comparable spatial resolution but without redshift information
(Salvador-Sol\'e \et 1993).

\begin{table*}[htbp]
\caption{Substructure analysis. Columns 6 is the highest value of
columns 3,4,5 and gives the final 
likelihood for the presence of substructure}
\begin{tabular}[l]{rlcccc}
\onerule                    
\multicolumn{1}{c}{Field}& 
\multicolumn{1}{c}{Name} &
\multicolumn{1}{c}{DS-test} & 
\multicolumn{1}{c}{Lee-test}&
\multicolumn{1}{c}{normality-tests} & 
\multicolumn{1}{c}{total}\\

\multicolumn{1}{c}{(1)} &
\multicolumn{1}{c}{(2)} &
\multicolumn{1}{c}{(3)} & 
\multicolumn{1}{c}{(4)} &
\multicolumn{1}{c}{(5)} & 
\multicolumn{1}{c}{(6)} \\
\onerule                    
   1& A~S~301   & 71& 38& 57& 71\% \\
   2& A~400     & 79& 99& 80& 99\% \\
   4& A~1060    & 41& 85& 99& 99\% \\
   5& A~S~639   & 94& 64& 75& 94\% \\
   6& A~3526    & 99& 96& 89& 99\% \\
   7& A~S~721   & 48& 13& 19& 48\% \\
   8& A~3556    & 81& 98& 78& 98\% \\
  9b& A~1736    & 99& 87& 34& 99\% \\
  11& A~3558    & 51& 28& 36& 51\% \\
 12a& A~S~805   & 98& 45&  9& 98\% \\
  13& A~3733    & 62& 22& 49& 62\% \\
  15& A~4038    & 90& 94& 93& 94\% \\
\onerule                    
\end{tabular}          
\end{table*}

\section{Segregation in velocity}

Radial velocities and magnitudes had to be normalized before a
kinematical analysis on the galaxy sample as a whole could
be started. We chose the most straightforward methods, using
absolute magnitudes $M$ and velocities normalized by cluster mean
velocity and velocity dispersion, i.e.~:
\begin{equation}
\tilde{v}_i = \frac{v_i - \langle v\rangle}{\sigma}
\end{equation}
This implies that $\tilde{v}_i$ has a mean value of 0 and a
standard deviation of 1 in each cluster. Normalized velocity
dispersions 
\begin{equation}
\sigma_{w} = \sqrt{\frac{\sum{(\tilde{v}_i)^2}}{N(N-1)}} 
\end{equation}
will be used in the following as kinematical indicators for different
samples. Only galaxies brighter than $M$~=~-19$^{mag}$ have been
included in the following analysis, which roughly corresponds to the
completeness limit of the present dataset. Choosing only objects
brighter than $M$~=~-19 also excludes dwarf galaxies, which could
contaminate fainter samples (Binggeli \et 1988).

\subsection{Dependence on galaxy types}
First, we looked at differences in the kinematical behaviour
of galaxies depending on their morphology. It can be seen in Figure 1
that there is a continuous trend of the velocity dispersion to increase
from early to late galaxy types.  Velocity dispersions have been
computed with a biweight estimator of scale (Beers \et 1990), which has
shown to be superior when only few objects are involved. Error bars
come from a bootstrapping calculation with 1000 iterations.  Between E
and S galaxy types a rise in the velocity dispersion of 30 \% can be
observed. The hypothesis that E--galaxies have the same velocity
dispersion as S0--galaxies cannot be excluded by an F--test (17 \%
likelihood). The same is true for the difference between S0 and
S--galaxies (23 \% likelihood of same underlying distribution), while
the difference between E and S--galaxies is significant at a level of 3
\%.

\begin{figure}[hbtp]
\centerline{\EPSFxsize=8.8cm \EPSFbox{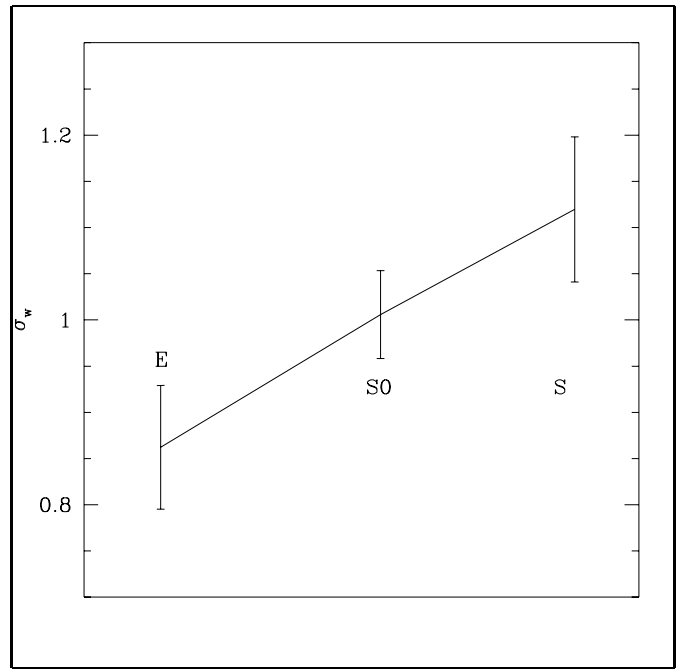}}
\caption[]{Velocity dispersion for galaxies of 
different morphological classes.}
\end{figure}

\subsection{Dependence on galaxy luminosity}
In the present sample of galaxy clusters there are clear signs of
velocity dispersion dependence on absolute magnitude for galaxies
brighter than $M \approx -22^{mag}$, as can be seen in Fig. 2. 
Again, velocity dispersions have been computed with a biweight
estimator of scale and error bars come from bootstrapping with
1000 repetitions. Bin limits were set every 0.5 mag between 
$M = -23.5^{mag}$ and $M = -19.0^{mag}$, taking the biweight mean value
$<M>$ in each bin as the x-position instead of the bin center.
Further dividing the
sample into galaxies of different types reveals that mainly the most
luminous early type galaxies are responsible for the
lower velocity dispersion. However, there is a general tendency of
ellipticals to have lower dispersions than spirals also at the faint
end. Former effect cannot be due exclusively to the existence of D/cD
galaxies residing in the bottom of the potential well, because only 4
of the clusters are given a Bautz--Morgan type I (Abell \et
1989). Another hint is that 
there are no E--galaxies with normalized velocities larger than
1.2 down to $-20.6^{mag}$, while 6 or 7 would be expected from a normal
distribution with $\sigma_w$ = 1.  No clear indication about the
kinematical status of S0 galaxies was found. It is widely known that 
there is considerable danger of confusion while classifying S0
galaxies as an intermediate class between E and S (Bender 1992). This
uncertainty in the morphological classification, together with the low
number of S0 galaxies brighter than $M = -22^{mag}$ in our sample,
makes it difficult to state about the presence of luminosity
segregation in this class. It seems that the brightest S0 galaxies lie
very close to the kinematical center of the cluster, indicating that
these galaxies have been residing in the centers of clusters for long
periods of time. On the other side, no signs of velocity dispersion
changes with luminosity can be seen for E or S0 galaxies fainter than
$M = -21.5^{mag}$. This can be explained by the fact that the
time-scale for dynamical friction exceeds the Hubble time when galaxies
of luminosity lower than L$_*$ are involved (Sarazin 1988).\\
As can be seen in Fig. 2, there is a deviating galaxy with $M \approx
-23.2^{mag}$ and $\sigma_w \approx 1.6$, which was classified as S by
Nilson (1973). Inspection of the corresponding photographic plate
reveals that the object is of peculiar nature, possibly interacting
with its neighbours. Its morphological type is given as uncertain in
several other catalogues, ranging between S0 and Scd. 

\begin{figure}[hbtp]
\centerline{\EPSFxsize=8.8cm \EPSFbox{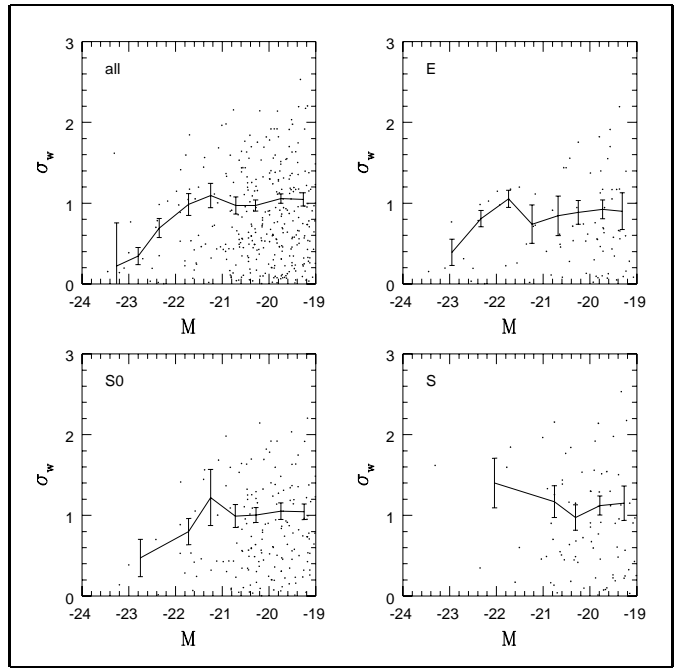}}
\caption[]{Velocity dispersions for galaxies of different
absolute magnitudes.}
\end{figure}

\section{Discussion}
We analysed the kinematics of the core regions of 12 nearby galaxy
clusters. The homogeneous redshift sample was nearly complete down to
faint limits ($\lapprox$ 18--19$^{mag}$) and was supplemented by
magnitudes and rough morphological types for most of the galaxies. An
analysis of substructure using velocity data revealed that 50 \%
of the cluster cores harbour significant substructure, thus confirming
that many clusters are not even relaxed in their inner regions, where
effects like mass segregation and infalling groups of galaxies might be
disturbing the virialization process.\\
After having merged the data to a sample of normalized galaxy
velocities, magnitudes and types, we looked for type and
type/luminosity segregation in velocity space.  Previous findings
(Binggeli \et 1987; Sodr\'e \et 1989) about early type galaxies having
lower velocity dispersions than late types are confirmed by the present
analysis. Zabludoff \& Franx (1993) found no such relation, on the
opposite they claimed deviations in the velocity means between
different types, concluding that there must be groups of spirals
falling onto the cluster main body and distorting the distribution of
velocities. Their findings should be considered complementary to ours,
because of the different scale observed (R $\leq$ 0.5 h$^{-1}$ Mpc
versus R $\lapprox$ 1.5 h$^{-1}$ Mpc).\\
Luminosity segregation in velocity space is also present, qualitatively
and quantitatively in agreement with the findings of Biviano \et
(1992), who were using a larger, but more heterogeneous data sample.
Moreover, there is a link between type and luminosity segregation. Only
the brightest E and, possibly, S0--galaxies ($M \lapprox$ -22$^{mag}$)
show clear signs of the phenomenon of luminosity segregation in
velocity, which is probably related to two--body relaxation
effects. These galaxies are responsible for the differences in velocity
dispersion between early and late galaxy types, plausibly representing
the fraction of galaxy population in clusters which have undergone
significant late dynamical evolution.  On the other side, S--galaxies
show no sign of luminosity segregation, as is expected from objects
that are still infalling onto the cluster main body and are presumably
crossing the core region for the first time. \\
It remains uncertain whether the effect is due to dynamical friction or
comes from the fact that galaxies which are the result of repeated
mergers will tend to have velocities closer to the cluster mean. In
both cases, strong evolution in the dynamics of the oldest cluster
population can be seen.  Capelato \et (1981) found several hints for
the presence of partial equipartition of galaxy kinetic energies in
clusters. If the ${\cal M}$/L ratio is close to constant for
E--galaxies, then our findings support their view that only the most
massive galaxies had had enough time to slow down.\\

\begin{acknowledgements}
The author thanks Dr. H.~MacGillivray, who made available COSMOS 
magnitudes for several clusters. I am specially indebted to
Dr. G.~A.~Tammann for his advice through all the stages of this project
and to Drs. E.~Salvador--Sol\'e and J.--M.~Solanes for valuable
comments on an original version of the manuscript. Detailed comments 
by the referee, Dr. A.~Mazure, helped improving the paper substantially. 
This research has made use of the Lyon-Meudon Extragalactic 
Database (LEDA, CRAL-Observatoire de Lyon) and of the NASA/IPAC 
Extragalactic Database (NED, JPL, Caltech). Support of 
the Swiss National Science Foundation is gratefully aknowledged.
\end{acknowledgements}


\end{document}